# Self-Channelling of Electric Current in a Quantum Well

S.A. Emelyanov[*]

*Division of Solid State Electronics, A.F. Ioffe Institute, 194021 St. Petersburg, Russia*

**Abstract**

We have directly demonstrated that homogeneous photoexcitation of a quantum well in presence of uniform tilted magnetic field gives rise to a *set* of bypass in-plane electric currents of a different value which may flow even *in the opposite directions simultaneously*. The effect has been observed in an asymmetric InAs quantum well under the Landau quantization. Theoretical model of the effect are discussed as well as the related problems.

## 1. Introduction

The effect of spontaneous self-channelling of electric current has not been observed up to now in any electronic systems. This seems to be quite natural because, as a minimum, the following hard requirements must be fulfilled in a system-candidate: (i) the electrons must be spatially separated and (ii) parameter(s) of their motion *along the same axis* must be a function of their coordinate along any other one. In fact, the situation which is very closed to the required one occurs in the well known system of 2D electrons under Landau quantization by transverse magnetic field. Indeed, according to the solution of the Schrödinger equation for such a system, the electrons' wave vectors along one axis in the well plane are simply proportional to their coordinate along another one (see, e.g., [1]). However, because of the axial symmetry of the system and, hence, degeneracy of the Landau levels related to the electrons' wave vector, their in-plane velocity is exactly zero and thus the above requirements are not fulfilled.

In present paper, we report on the first observation of a self-channelling of electric current in a 2D system. The effect has been observed in an asymmetric InAs quantum well under the Landau quantization by tilted magnetic field. The current has been induced by far-infrared laser radiation.

## 2. Experimental

The samples studied consist of a single layer of InAs sandwiched between a thick GaSb buffer layer and 20 nm wide GaSb cap layer. Since the system has a broken gap alignment, i.e. conduction band edge of InAs lies 150 meV below the valence band of GaSb, there were two thin (3 nm wide) AlSb barriers surround the InAs layer to avoid any hybridization-related effects. The samples were grown by MBE method so that the electrons in the well may be created either by donor-like surface states [2,3], or by donor-like interface states [4]. Taking into account that in our structures the electron sheet density is about $1.6\times10^{12}$ cm$^{-2}$ (mobility $1.5\times10^5$ cm$^2$/Vs) while the number of electrons from the interface states is estimated to be about $5\times10^{11}$ cm$^{-2}$ [5], one can suppose the majority of electrons to be caused by the surface states. This source of electrons is obviously asymmetrical, i.e. it gives rise to a strong transverse electric field in the well, pointing toward the substrate side. Schematically, the energy-band diagram is shown in the left-hand panel of Fig. 1. Assuming such a "built-in" field to be uniform one can estimate its value to be of the order of $10^5$ V/cm (see, e.g., [5]) but, in fact, local value of the field may be even higher because the potential profile should be rather exponential than linear. The samples (typically $8\times10$ mm) are supplied by indium ohmic contacts. Geometry of the contacting will be discussed below.

---

[*] E-mail: sergey.emelyanov@mail.ioffe.ru

We study the in-plane currents induced by far-infrared (FIR) radiation from pulsed $NH_3$ laser optically pumped by $CO_2$ laser. The wavelength is 90.6 μm ($\hbar\omega = 13.7$ meV), pulse duration is about 40 ns, maximum radiation intensity $I_m \approx 200$ W/cm$^2$. In most of the experiments the radiation is linearly polarized. The shape of the exciting laser pulses is monitored by using of high-speed photon-drag detector. Typical dynamics of the laser pulse is shown in the right-hand panel of Fig. 1. In-plane current pulses in *unbiased* structures are detected by digital storage oscilloscope (Tektronix TDS 784D) through the voltage drop on a 50 Ohm load resistor in a short-circuit regime. The incident light is normal to the sample surface. The external magnetic field $B$ is tilted from the normal on the angle $\alpha$. Most of the experiments have been performed at $T = 2.2K$.

**3. Results**

Pronounced FIR-light-induced current pulses have been observed in the samples supplied by a pair of strip-like ohmic contacts. Geometry of the experiment as well as geometry of the contacting are shown in the inset of Fig. 2. The shape of the current pulses is similar to that of the laser pulses indicating that response time of the effect is less than $10^{-8}$ s. The sign of the current is independent of the sign of $B_z$ and it reverses with reversing of the sign of $B_x$. At $B_x = 0$ the current disappears. For symmetry reasons, the observed current should be the result of the breaking of axial symmetry of the system. The presence of only in-plane component of magnetic field is clearly not enough. As a minimum, there must be also a polar vector directed across the well. The origin of such vector has become evident from the following experiments. If we take the structures in which AlSb barrier in the right side of the well (see diagram in Fig. 1) is as thick as 100 nm, then the current decreases drastically and, moreover, in some structures its sign may be even reversed. The similar result has been observed on the structures in which GaSb cap layer is as thick as 120 nm. These facts show that the vector sought is just the built-in electric field provided by the surface donor states. Indeed, it is well known that AlSb thick layers may be the source of electrons for InAs quantum well owing to a deep-donor defect in the bulk of AlSb [6]. The deep-donor-related field may compensate the surface-related one, making the InAs well to be more symmetric. Hence, the uncontrolled interface-donor-related field may become of a major importance. The similar result can be obtained when the well-to-surface distance is increased by incorporation of a thick GaSb cap layer. Finally, we study several samples from the same wafer which are oriented along different crystallographic axis. We find no remarkable difference in the behaviour of the light-induced current in these samples that means the effect of bulk asymmetry of the host material on the current is of a minor importance.

Fig. 2 shows the behaviour of the current along $Y$ axis ($J_y$) as a function of magnetic field. It is seen that there are two pronounced branches. In the low-field one ($B < 2$ T), the current can easily be identified as a photogalvanic current (see, e.g., [7]) which was observed earlier in asymmetric 2D systems [8-9]. At higher $B$, the effect disappears that is quite natural because of the Landau-quantization-related drop of ohmic conductivity. Surprisingly, at $B > 3$ T the current appears ones again (high-field branch) showing a pronounced resonant behaviour. Position of the resonance correlates with that of the cyclotron resonance (CR). Although corresponding effective mass ($m^* \approx 0.04 m_0$) is slightly higher than the commonly used value (about $0.033 m_0$) this could be explained, for example, by variation of $m^*$ as a function of Fermi energy [10]. Identification of the resonance at the high-field branch as to be CR-related is confirmed also by the experiments in which the exciting light was circularly polarized: under CR-active polarization the resonance remains while under CR-inactive one it disappears within experimental error. The fact that the photocurrent resonance is relatively wide could be explained by a saturation effect. Indeed, following to Ref. [11], reduction of the absorption coefficient in a factor of 2 may occur in our spectral range at as low intensity as about 30 W/cm$^2$ on the samples in which the electron density is only in a factor of 3 lower than that in our samples. The only principle difference in the behaviour of the current spectra with respect to transmission ones is the presence of a characteristic dip shifted from the resonant point toward the low magnetic fields. This is despite of the fact that asymmetric InAs quantum wells in tilted quantizing magnetic fields yield a symmetrical CR line in transmission spectra [12].

The quite unexpected (and even paradoxical) result we have obtained studying the current in the same geometry as in the inset of Fig. 2 but through two pairs of short ohmic contacts. Geometry of the contacting is shown in the inset of Fig. 3(a). The length of the contacts are about one fifth of the shorter side of the sample. They are symmetrical with respect to the line $x = W/2$ ($W$ is sample width) and shifted toward the opposite borders of the sample. Under homogeneous excitation of whole sample we measure the current through each pair (1-2 and 3-4) for two opposite directions of magnetic field ($B^+$ and $B^-$). At first sight, one would expect that the currents will flow in the same direction as that in Fig. 2 but will be lower, roughly, in the rate of the arias covered by the contacts. However, the real picture differs in principle from the expected one. Figs. 3(a) and 3(b) show the behaviour of the currents through each pair of contacts at $B = B^+$ and at $B = B^-$, respectively. It is seen that the currents flow *in opposite directions simultaneously* while the sample should be quite homogeneous in the well plane. Moreover, the current $J_y^{1-2}$ is even higher than the current from the whole sample (see Fig. 2). To exclude any effects related to an accidental non-homogeneity we repeat the experiment after the rotation of the sample in $XY$ plane on the angle $180°$. Distribution of the currents remains. It is also seen from Figs. 3(a) and 3(b) that switching of $B$ from $B^+$ to $B^-$ does not result in the reverse of the sign of both $J_y^{1-2}$ and $J_y^{3-4}$. Instead, it leads to a reversed mirror reflection of the currents with respect to the line $x = W/2$. Furthermore, *position* of the resonances is also sensitive to the contacting: on both figures the lower peak shifts toward the lower magnetic fields while the higher one shifts toward the higher fields. This fact clearly correlates with the behaviour of the spectra in Fig. 2. Indeed, the maximum of the lower current corresponds with the characteristic dip on these spectra whereas the maximum of the higher one corresponds with the hump. Once more, rotation of the sample on the angle $180°$ in $XY$ plane does not change either position or amplitude of the resonances.

Finally, to make the results obtained to be unambiguous, we have carried out a one more qualitative experiments. We etch away those fragments of the well which are not covered by the contacts and then put the sample back at the same position with respect to the laser beam (Fig. 4). As expected, the sign of the currents through each pair is identical to the sign of the current shown in Fig. 2 while their value is roughly one fifth of the current in Fig. 2.

### 4. Discussion

It is understood that detailed theoretical explanation of the observed effects is a separate problem and we do not claim to resolve it in present work. Nevertheless, we would like to discuss several ideas which could be a starting-point for further investigations. First, let us consider the solution of Schrödinger equation for two coupled $\delta$-shaped quantum wells remote from each other on a distance $a$ in a parallel magnetic field $B_x$. Following to Ref. [13], here $k_y$-dependent correction appears to the energy of Landau levels. For zeroth Landau level, this correction is:

$$E'(k_y) \equiv E'(z_0) = -\frac{1}{\sqrt{\pi}\lambda}\left(\Delta_1 \exp-\left(\frac{z_0 + a/2}{\lambda}\right)^2 + \Delta_2 \exp-\left(\frac{z_0 - a/2}{\lambda}\right)^2\right), \quad (1)$$

where $\Delta_{1,2} = \int V_{1,2}(z)dz$; $\lambda$ is magnetic length; $z_0 = k_y\lambda^2$ is position of the center of electron cyclotron orbit. It is seen from Eq. (1) that in such a system the electrons remain to be spatially separated depending on their wave vector ($z_0 = k_y\lambda^2$) like in the case of conventional 2D system in purely transverse magnetic field. However, in our system the Landau level degeneracy related to the electrons' wave vector is lifted so that the electrons' velocity along $Y$ axis ($v_y \propto dE(k_y)/dk_y$) is nonzero and it is a function of electrons' position along Z axis ($z_0$). Note also that if potential profile is asymmetric ($\Delta_1 \neq \Delta_2$) then the energy spectrum is asymmetric as well: $E(k_y) \neq E(-k_y)$. In fact, the above solution implies the electron gas to be a set of spatially separated 1D channels along Y axis in which the electrons flow in opposite directions remaining the net flux to be equal to zero. As it has also been shown in [13], in such system the electrons possess an electric dipole moment along Z axis correlated with their velocity along Y axis that can be considered as just the result of the Lorentz force effect on the electrons in the channels. Accordingly, the net polarization of undisturbed system is also equal to zero.

The advantage of the model discussed is that it allows one to get a transparent analytical solution of the Schrödinger equation. However, it is clearly very idealised since potential profile is supposed to be $\delta$-shaped. As a result, the size-quantization effect along Z axis is ignored while it is known to be important for any real 2D systems. Unfortunately, Schrödinger equation has not been solved for the more realistic situation of an arbitrary potential profile $V(z)$ when both in-plane ($B_x$) and quantizing ($B_z$) components of the magnetic field are nonzero. Nevertheless, it has been shown analytically that in such system the electrons may also possess nonzero electric dipole moment along both Z and X axis as a result of the presence of local currents along Y axis [14]. As a minimum, this allows one to believe that the model of $\delta$-shaped potentials is not so fantastic and may be relevant to real 2D systems. On the other hand, it is easy to see that the model of spatially separated electrons possessing nonzero velocity along Y axis implies that the optical transitions between Landau levels may result in a set of light-induced electric currents along Y axis flowing in the opposite directions. It is only necessary to assume that (i) the energy spectrum of the system is asymmetrical in k-space and (ii) electrons' velocity and, hence, their dipole moment is a function of the Landau quantum number. Both assumptions are in agreement with theoretical calculations of Ref. [14]. Then, it seems to be quite reasonable to assume that the optical transitions between the states with different velocities result in a set of light-induced currents along Y axis which are not compensated each other due to their spatial separation. The characteristic radius of each current channel is thus of the order of radius of the electron cyclotron orbit, i.e. it is a function of the external magnetic field.

It is also important that the approach proposed allows one to explain qualitatively the observed shift of CR peak position as a function of coordinate along X axis. To do that we use an analogy despite of the fact that applicability of any analogy is always an open question and requires further theoretical confirmation. The point is that a 2D system which consists of a number of oriented dipoles is known in physics: it is the so-called Langmuir-Blodgett films which consist of a number of oriented molecular dipoles (see, e.g., [15] and refs. therein). An external transverse electric field, which is approximately of the same order as that in our system, is known to result in a shift of the energy of optical transitions between ground and excited states of the molecule (Shtark effect). For the linear Shtark effect this shift is: $\Delta \hbar \omega = \Delta d E \cos \Theta$, where $\Delta d$ is changing of the dipole moment caused by the optical transition; $\Theta$ is the angle between the electric field and the dipole axis. By analogy, for our system this formula can be written as: $\Delta \hbar \omega = (d_{N+1} - d_N) E \sin \alpha$, where $d_N$ and $d_{N+1}$ is electron's dipole moment at initial and final state of the optical transition, respectively. It is important that the sign of $\sin \alpha$ depends on the sign of electrons' velocity on the Landau level (see inset of Fig. 3(b)). Therefore, if the electrons on the Landau levels truly flow in opposite directions at fixed magnetic field, then CR peak should split into a set of neighbouring CR peaks the position of which may be shifted either at higher or at lower B. This should result (i) in an asymmetry of CR peak under the strip-like contacting and (ii) in a shift of CR peak position for the currents through different fragments of the sample. This is just what we observe in the experiment. It is seen that, according to the proposed simplest model the effects discussed should be more pronounced at higher tilt angles. To test this point, we have studied current spectrum at higher tilt angle: $\alpha = 20°$. The results are shown in Fig. 2 (curve 3). Indeed, CR peak asymmetry has clearly become more pronounced and, moreover, the distance between the characteristic dip and hump on the curves in Fig. 2 increases roughly in the rate of $\sin \alpha$.

In the context of developed ideas the important question is whether there is a current along $X$ axis and what about its behaviour, if any. To answer this question, we have carried out the experiments in the geometry shown in the inset of Fig. 5. The results are as follows. When the light intensity is as low as about 5W/cm$^2$, we observe a bipolar photo-response under CR conditions but, at higher $I$, it tends to transform into a unipolar one. Evolution of dynamics of the photo-response along $X$ axis with increasing of the laser intensity is shown in Fig. 6. In principle, the low-intensity photo-response (Fig. 6(a)) looks like a first derivative of the exciting laser pulse and, hence, could be interpreted as a non-steady displacement current resulting from an imbalance in the electrons' dipole moments under photoexcitation. However, as an alternative, one would suppose such a photo-response to be the result of a difference in relaxation times for the currents flowing in opposite directions. Since the relaxation process should be very sensitive to the lattice temperature we have studied dynamics of photo-response at the temperature reduced down to 1.8K. The result is that we have not observed any changing in this dynamics that is an argument in favour of the former mechanism. Instead, we have observed that at lower temperatures transformation of the shape of photo-response from bipolar to unipolar occurs at higher laser intensities. This remarkable fact will be discussed below.

Of course, the most surprising result is the emergence of a unipolar current along $X$ axis. At first sight, this seems to be a failure of proposed model, especially, because of the behaviour of the current $J_x$ as a function of $B$ at different angles $\alpha$ (Fig. 5) is reminiscent that of the current $J_y$ (see Fig. 2). Moreover, under CR conditions, $J_x$ is clearly even higher than its counterpart $J_y$. To resolve this puzzle, we have carried out a series of qualitative experiments using the method of partial exposition of the sample to the laser light. We have used a non-transparent plate with a rectangular window the length of which is equal to the sample length while the width is about one fifth of that of the sample. First, we measure the current $J_y$ when the only two nearest to the borders parts of the sample are exposed to light (Fig. 7(a)). We have find that these currents flow in the opposite directions that is clearly consistent with our main result (see Fig. 3). After that, we rotate the sample together with the window on the angle 90° about $Z$ axis and then repeat the experiment (Fig. 7(b)). Here the current is found to be independent of the position of the window. Finally, we carry out a mixed experiment: the sample is oriented along $X$ axis while the window is oriented along $Y$ axis (Fig. 7(c)). The result is similar to that observed in the first experiment: the current is a function of the window position that is also consistent with its behaviour in Fig. 5. Thus, summarising the results of these experiments, one can come to the conclusion that (i) self-channelling of electric current occurs only along $Y$ axis but not along $X$ axis and (ii) despite of their similar behaviour, the currents $J_y$ and $J_x$ are of a different nature.

Clarification of the nature of the unipolar current $J_x$ comes from the experiments with a circular polarized light. As it was mentioned above, the current $J_y$ occurs only at CR-active light polarization. By contrast, the current $J_x$ is found to be caused by CR-inactive polarization as well as by CR-active one and, moreover, in the former case it is even somewhat higher. This fact shows that these currents are caused by *different optical transitions* which obey the different selection rules. We see no other candidate on the role of alternative optical transitions than the so-called indirect ones. These transitions are accompanied by scattering on acoustic phonons so that the electrons' wave vector is not conserved. Although such an assumption seems to be too much exotic, it, nevertheless, allows one to explain the number of experimental facts. The first one is the above difference in selection rules for $J_y$ and $J_x$. The second one is reduction of the unipolar $J_x$ with respect to the bipolar one with decreasing of the lattice temperature. The third one is the relative increasing of the unipolar $J_x$ at high laser intensities, i.e. its insensitivity to the saturation effect. Furthermore, such an unexpectedly high contribution of indirect optical transitions in the current along *X* axis is also consistent with proposed model. Indeed, if the self-channelling is truly related to the spatial separation of electrons along *X* axis depending on their wave vector along *Y* axis, then the indirect optical transitions must be accompanied by a hopping of electrons in real space along *X* axis. The asymmetry of the energy spectrum should result in an imbalance of the hopping along *X* axis and, hence, in a nonzero net hopping. Intuitively, the relative efficiency of such mechanism seems to be very high but, of course, theoretical confirmation is required in order to legitimate it.

Finally, a separate problem we would like to discuss briefly in the context of presented experimental results. The point is that if the proposed physical picture will prove to be valid, then it follows immediately that spontaneous 1D electric currents exist even within the fully occupied Landau levels, i.e. even at the ground state of the system. In this sense, the system may be considered as to be in a specific phase in which macroscopic 1D electron fluxes flow in opposite directions along some axis. Such a flux phase clearly differs conceptually from any known ones which imply a staggered-vorticity behaviour of the fluxes on the scale of few angstroms (see, e.g., [16-17]). In itself, the existence of such phase may seem to be impossible or, at least, to be quite paradoxical. Nevertheless, the number of presented experimental facts, which could be explained by using this concept, make it to be worthy of further detailed study.

**Summary**

We have directly demonstrated that homogeneous photoexcitation of a quantum well in presence of tilted quantizing magnetic field gives rise to a set of bypass in-plane electric currents of a different value which may flow even in the opposite directions. Such a self-channelling of electric current has been observed in an asymmetric InAs quantum well under the Landau quantization. Various aspects of this phenomenon have been studied experimentally. Theoretical model has been discussed together with the related problems.

**Acknowledgements**

.
Gratefully acknowledged are B.Ya. Meltser and S.V. Ivanov (Ioffe Institute) for the MBE samples, Yu.V. Kopaev for useful discussion.

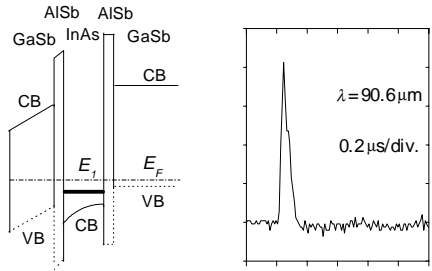

FIG. 1. Left-hand panel: schematic energy-band diagram of the system studied. Right-hand panel: typical track of the exciting laser pulse.

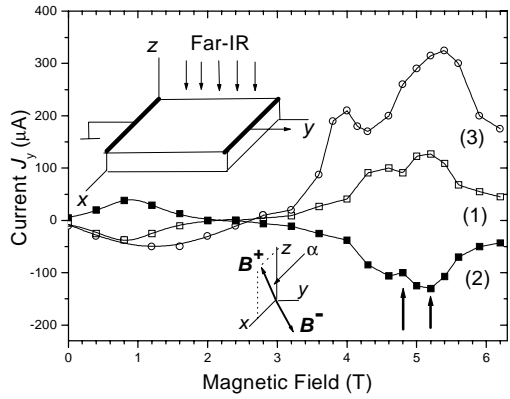

FIG. 2. FIR-light-induced current along *Y* axis as a function of magnetic field ( $I = 200W/cm^2$, $T = 2.2K$ ) : (1) - $B = B^+$, $\alpha = 8°$; (2) - $B = B^-$, $\alpha = 8°$; (3) - $B = B^+$, $\alpha = 20°$. Solid lines are a guide for the eyes. Vertical arrows show position of characteristic dip and hump on the curves (1) and (2). The inset shows geometry of the experiment.

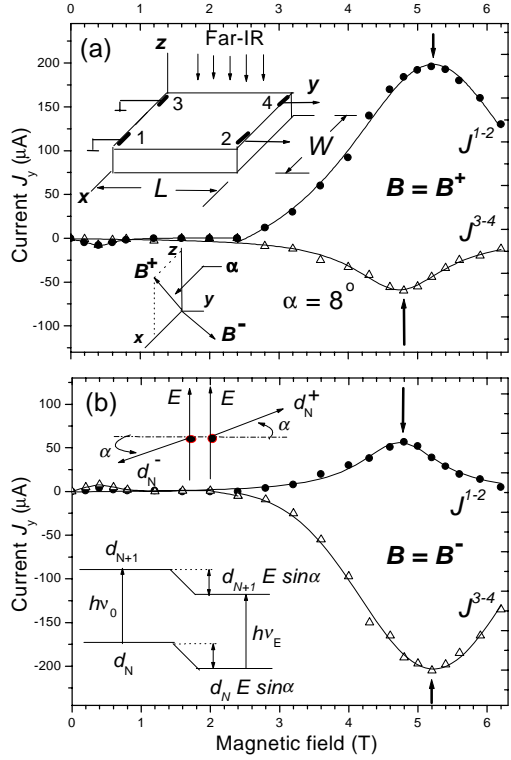

FIG. 3(a) FIR-light-induced current through the contacts 1-2 and 3-4 at $B = B^+$ and $\alpha = 8°$ ($I = 200 W/cm^2$, $T = 2.2 K$). The lengths of the contacts are about $W/5$. They are positioned symmetrically with respect to the line $x = W/2$ and shifted toward the opposite sample borders. Vertical arrows show CR peak position. The inset shows geometry of the experiment.
(b) FIR-light-induced current through the contacts 1-2 and 3-4 under the same conditions as in the upper panel but under reversed magnetic field: $B = B^-$. Vertical arrows show CR peak position. The inset illustrates possible mechanism of the shifting of CR peak position related to the opposite orientation of the dipole moments of electrons on the Landau levels.

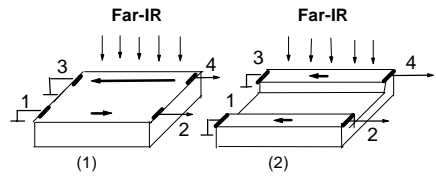

Fig. 4. Schematic diagram of distribution of the currents along $Y$ axis under the same conditions as in Fig. 3(b): (1) – before etching; (2) – after etching of those fragments of the sample which are not covered by the contacts. The length of thick arrows is proportional to the value of the currents.

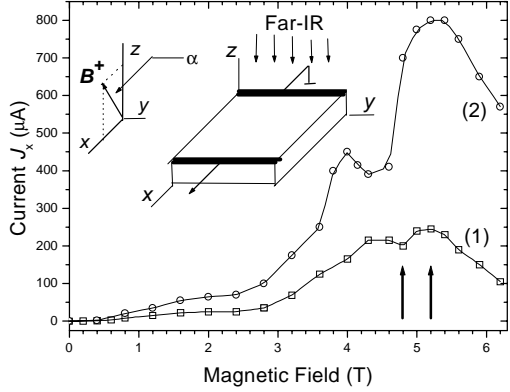

FIG. 5. FIR-light-induced current as a function of magnetic field under the same conditions as in Fig. 2 but the sample is rotated on the angle $90°$ about $Z$ axis: (1) - $B = B^+$, $\alpha = 8°$; (2) - $B = B^+$, $\alpha = 20°$. Solid lines are a guide for the eyes. Vertical arrows show the position of characteristic dip and hump on the curve (1). The inset shows geometry of the experiment.

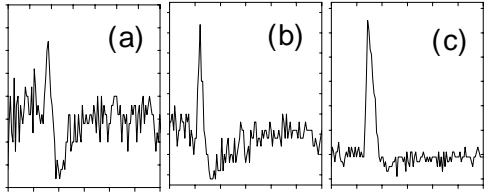

FIG. 6. Evolution of dynamics of the photo-response in the experimental geometry of Fig. 5: (a) - $I = 5$ W/cm$^2$; (b) – 15 W/cm$^2$; (c) – 50 W/cm$^2$. The time scale is $0.2 \mu s$ /div.

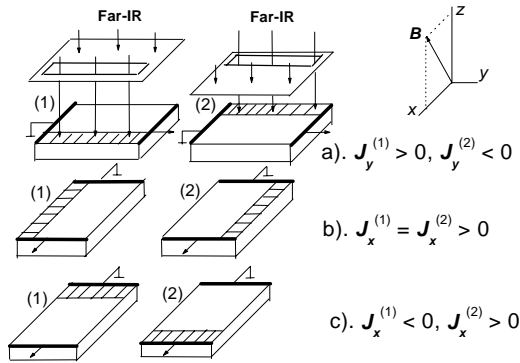

FIG. 7. Sketch of experiments using the method of partial exposition of the sample to FIR light. Shading shows the fragments exposed. The results are shown qualitatively for three different orientation of both the sample and the window with respect to the magnetic field.